\documentclass[%
 reprint, amsmath,amssymb, physrev,superscriptaddress
]{revtex4-2}

\usepackage{graphicx}
\usepackage{dcolumn}
\usepackage{bm}
\usepackage{hyperref}
\begin{document}

\title{Quasi-monoenergetic multi-GeV electron acceleration in a plasma waveguide}

\author{Ronan Lahaye}
\affiliation{Laboratoire d’Optique Appliqu\'ee, ENSTA, CNRS, Ecole polytechnique, Institut Polytechnique de Paris, 828 Bd des Mar\'echaux, 91762 Palaiseau, France}
\author{Kosta Oubrerie}
\affiliation{Universit\'e Paris-Saclay, CEA, LIDYL, 91191 Gif-sur-Yvette, France}
\author{Olena Kononenko}
\author{Julien Gautier}
\author{Igor A.  Andriyash}
\author{Cedric Thaury}
\email{cedric.thaury@ensta-paris.fr}
\affiliation{Laboratoire d’Optique Appliqu\'ee, ENSTA, CNRS, Ecole polytechnique, Institut Polytechnique de Paris, 828 Bd des Mar\'echaux, 91762 Palaiseau, France}


\begin{abstract}

Laser-plasma accelerators present a promising alternative to conventional accelerators. To fully exploit the extreme amplitudes of the plasma fields and produce high-quality beams, precise control over electron injection into the accelerating structure is required, along with effective laser pulse guiding to extend the acceleration length. Recent studies have demonstrated efficient guiding and acceleration using hydrodynamic optically field-ionized (OFI) plasma channels. This guiding technique has also been combined with controlled electron injection to produce high-quality electron beams at the GeV level using a 50 TW laser. The present work extends these results to higher laser energies, demonstrating the generation of quasi-monoenergetic electron beams with peak energies exceeding 2 GeV, for a PW-class laser.

\end{abstract}

\maketitle

\section{Introduction}
Laser-wakefield acceleration relies on the use of an ultra-short 
and ultra-intense ($I\gtrsim10^{18}$~W.cm$^{-2}$) laser pulse, ionizing a light gas, and creating an underdense plasma. Propagating through the plasma, the laser pulse drives an accelerating structure for the electrons in its wake~\cite{PhysRevLett.43.267}. To maintain this accelerating structure over long distances, it is necessary to keep the laser pulse focused along the propagation in the plasma. One way to achieve this is to create a long plasma filament beforehand and let it expand radially to obtain a curved radial density profile with a minimum electron density on the laser axis. This density profile corresponds to a curved refractive index profile, with a maximum at the center, acting then as graded-index optical fiber, keeping the driver pulse focused along the propagation. Several energy records in laser-plasma acceleration were established using guides created with capillary discharges~\cite{2006NatPh...2..696L,PhysRevLett.122.084801}. However, this method makes it difficult to tailor the longitudinal density profile and is especially prone to material damage. 

These difficulties have led to the development of an alternate guiding scheme, using a secondary laser pulse focused in a line by an axicon~\cite{PhysRevAccelBeams.22.041302,PhysRevE.102.053201,PhysRevLett.125.074801} or an axiparabola~\cite{Smartsev:19,Oubrerie_2022b}, creating a quasi-Bessel beam with a long focal line over which the laser spot size varies slowly. 
This secondary laser pulse  generates, through optical-field ionization, the plasma filament which ultimately results in the formation of the waveguide. 
This technique was first demonstrated using density transition injection~\cite{PhysRevLett.86.1011} with 100 TW-class lasers, allowing the reliable production of high-quality electron beams at the GeV energy level~\cite{Oubrerie2022}. The first attempt with a PW laser, but without controlled injection, produced beams of up to 5 GeV, but with a large shot-to-shot variation, and a significant component at low energies~\cite{10.1063/5.0097214,PhysRevX.12.031038}.
Later, in another study, a 100 TW-class laser was used to generate high-quality GeV electron beams, confirming the robustness of controlled injection at this power level~\cite{PhysRevLett.131.245001}. Finally, the implementation of controlled injection with a PW laser enabled the generation of high-quality multi-GeV beams~\cite{PhysRevLett.133.255001}.

Here, we build upon these works employing the shock-assisted ionization injection technique~\cite{2015NatSR...516310T} to produce controlled multi-GeV electron beams with peaked energy spectra.


 
\section{Experimental setup}\label{sec:rules_submission}

\begin{figure}
    \centering
    \includegraphics[width=\linewidth]{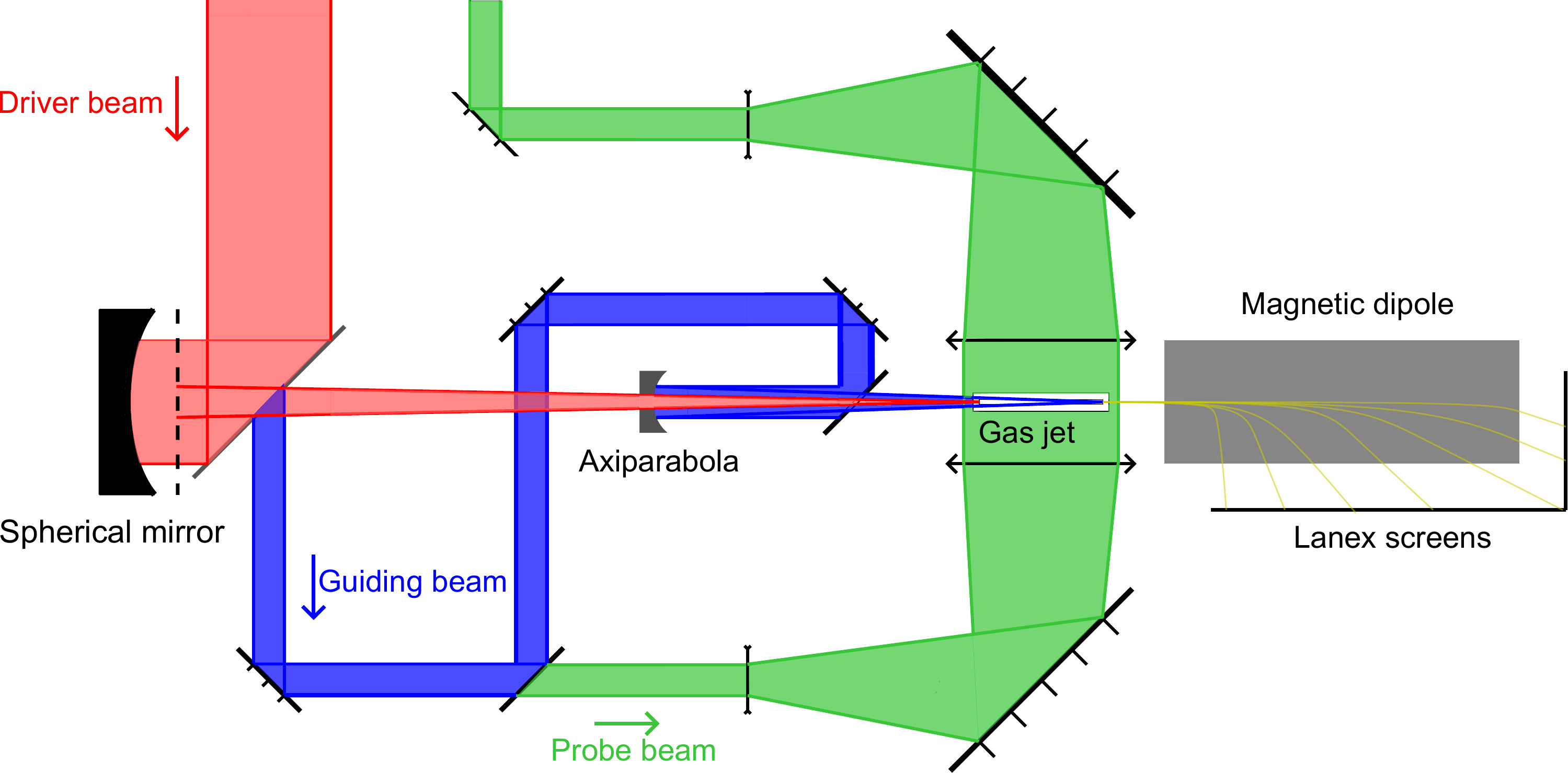}
    \caption{Simplified setup of the experiment. The beam generating the waveguide is represented in blue, the probe pulse in green, and the pulse generating the wakefield in red. The electron energy is  measured by a spectrometer composed of a dipole magnet and a LANEX scintillating screen. The long focal distance of the spherical mirror is not shown and represented by the horizontal dotted line for simplicity.}
    \label{fig:setup}
\end{figure}

\begin{figure}
    \centering
    \includegraphics[width=\linewidth]{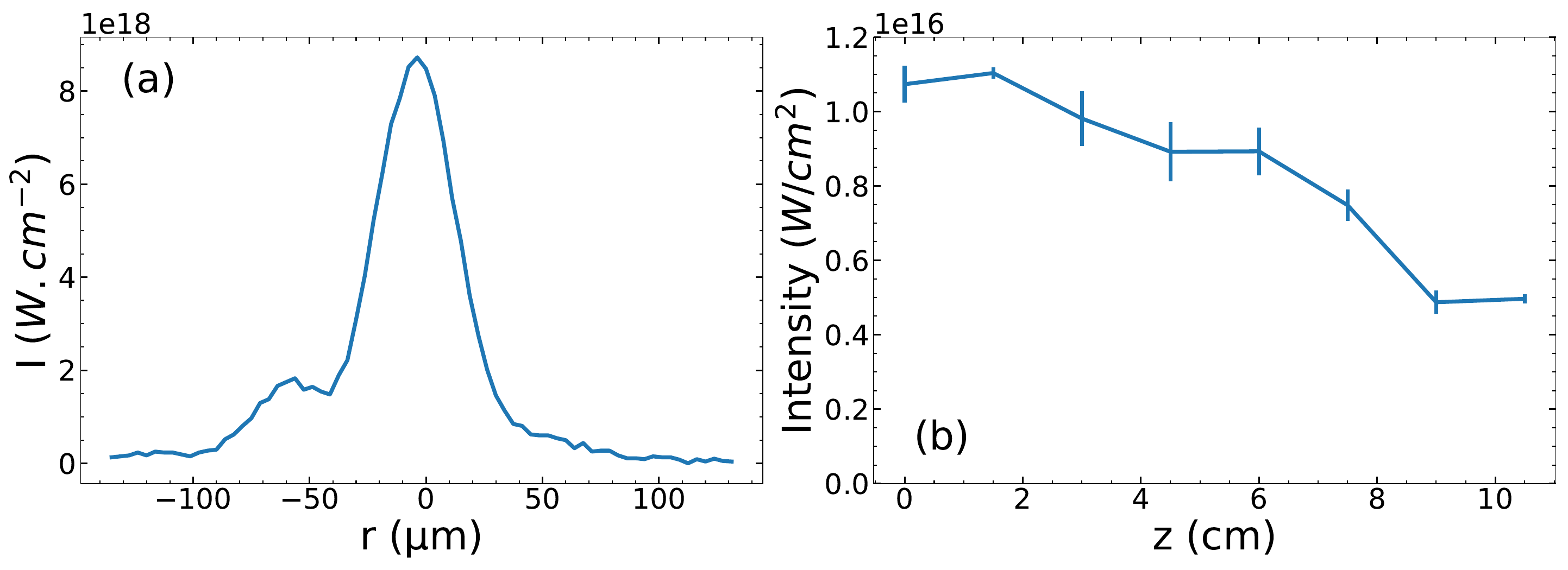}
    \caption{Focal spot of the driver beam and evolution of the intensity along the focal line of the guiding beam. (a) Measured focal spot of the driver. (b) Evolution of the intensity on-axis of the guiding beam, the error bars represent the statistical variation over 5 shots.}
    \label{fig:intens_P1_P2}
\end{figure}

The experiment was carried out at the Apollon laser facility in the Long Focal Area (LFA), using the F2 beam which delivers up to 12\,J, in a 30\,fs laser pulse at a central wavelength $\lambda_0=800$~nm . A simplified setup of the experiment is displayed in Fig.~\ref{fig:setup}.
The beam is focused by a 6-meter focal-length spherical mirror to a 45~$\mu$m full-width-at-half-maximum (FWHM) focal spot, as shown in Fig. \ref{fig:intens_P1_P2}(a), resulting in a normalized potential vector $a_0\approx2$. The guiding pulse is generated by a portion of the main pulse extracted through a holed mirror before the spherical mirror which focuses the main beam. It is attenuated down to $\approx$ 40\,mJ, then focused by an axiparabola to create a  focal line about 8\,cm long which generates a plasma filament in the gas target. The evolution of the on-axis intensity is shown in Fig. \ref{fig:intens_P1_P2}(b). The filament then expands for 6\,ns before the main pulse is focused in the guide and generates the wakefield. The delay between the guiding pulse and the driver pulse is adjusted through various delay lines in the chamber not represented in Fig.~\ref{fig:setup}. 


\begin{figure}
    \centering
    \includegraphics[width=\linewidth]{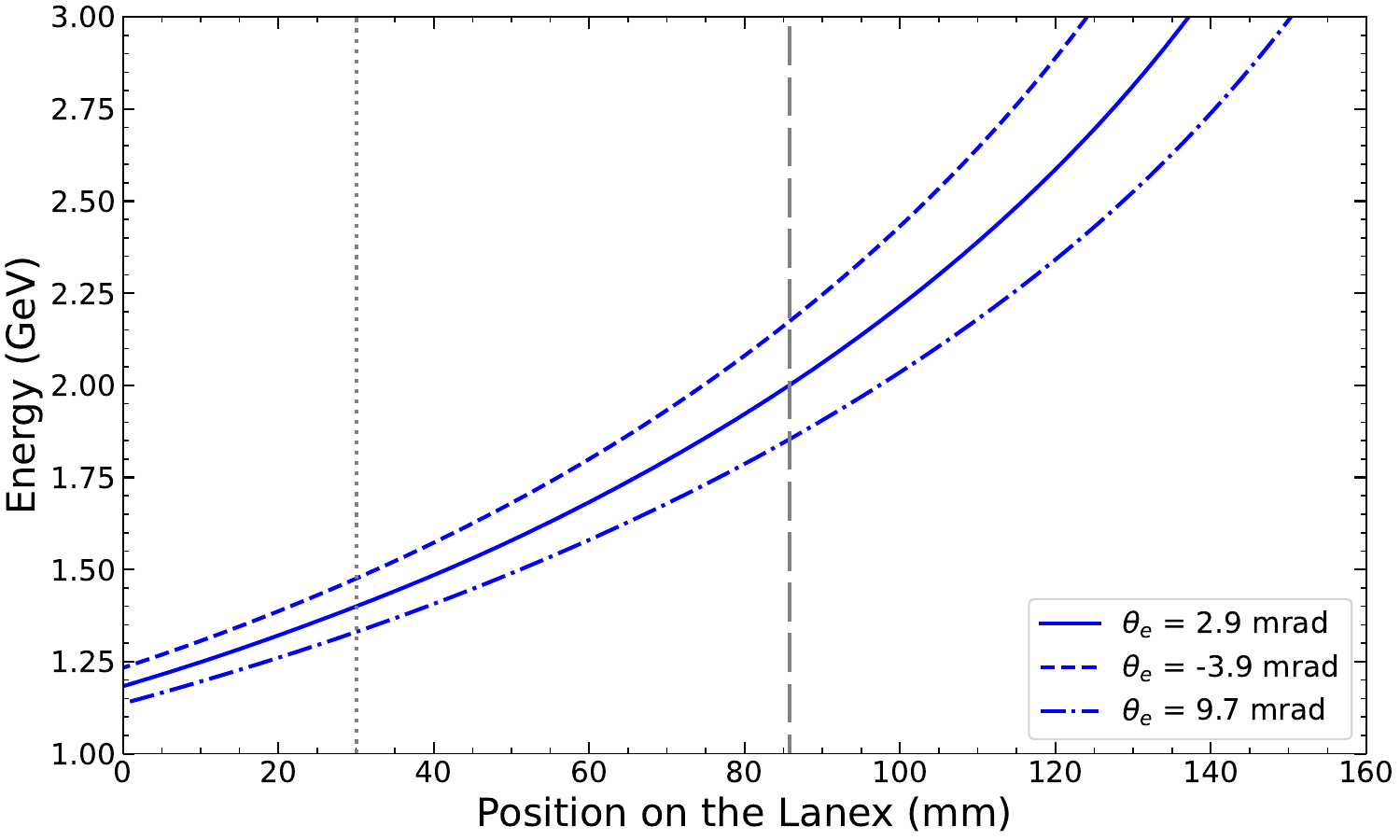}
    \caption{Dispersion relation of the electron spectrometer. Three dispersion relations are shown for three different entrance angles in the dipole. The vertical lines correspond to the position of an electron beam at 1.4 and 2~GeV respectively for $\theta_e=2.9$~mrad.}
    \label{fig:comp_dispersion}
\end{figure}

The target is a 6~cm long slit nozzle supplied with a mixture of 1\% Nitrogen and 99\% Helium gas at a backing pressure of 60\,bars. We estimated the average neutral density along the propagation axis by measuring the phase shift induced by the propagation of a low-power laser beam into the gas flow with a wavefront sensor. Using this technique, we measure an average density of $n_e\sim1\times10^{18}~\text{cm}^{-3}$, 2~mm above the jet and for an operating pressure of 60~bar. 
A small blade can be placed at the entrance of the nozzle to tailor the longitudinal density profile and trigger shock-assisted ionization injection, typically producing low-charge and low-dispersion electron beams~\cite{2015NatSR...516310T}.
A  120~cm permanent magnetic dipole deflects the electron beam which then impacts on scintillating Lanex Regular screens. The screens are imaged onto several CMOS cameras, and the resulting signal is analyzed to reconstruct the energy spectrum of the electron beam. A 180 $\mu$m thick magnesium membrane isolates the dipole vacuum chamber from the main experimental chamber. The crossing of this membrane slightly increases the electron beam divergence. In a separate experimental campaign, under similar conditions, the bremsstrahlung radiation produced by the crossing of the membrane was recorded, allowing to measure the entrance angle of the electron beam into the magnet for each shot. During our experiment, we did not have this diagnostic, so we used this data to compute three different dispersion relations, corresponding to an angle of the electron beam and the magnet axis of $\theta_e=2.9\pm6.9$~mrad, where a positive angle corresponds to the counterclockwise direction, and the uncertainty corresponds to the standard deviation of the distribution. The dispersion relations obtained are shown in Fig. \ref{fig:comp_dispersion}. Considering an electron beam at 2~GeV for an angle of 2.9~mrad, the uncertainty due to the variation of the angle of the electron is about 160~MeV, while it is about 72~MeV for an electron beam at 1.4~GeV for the same angle.

\section{Results}\label{sec:types_paper}

\begin{figure}
    \centering
    \includegraphics[width=\linewidth]{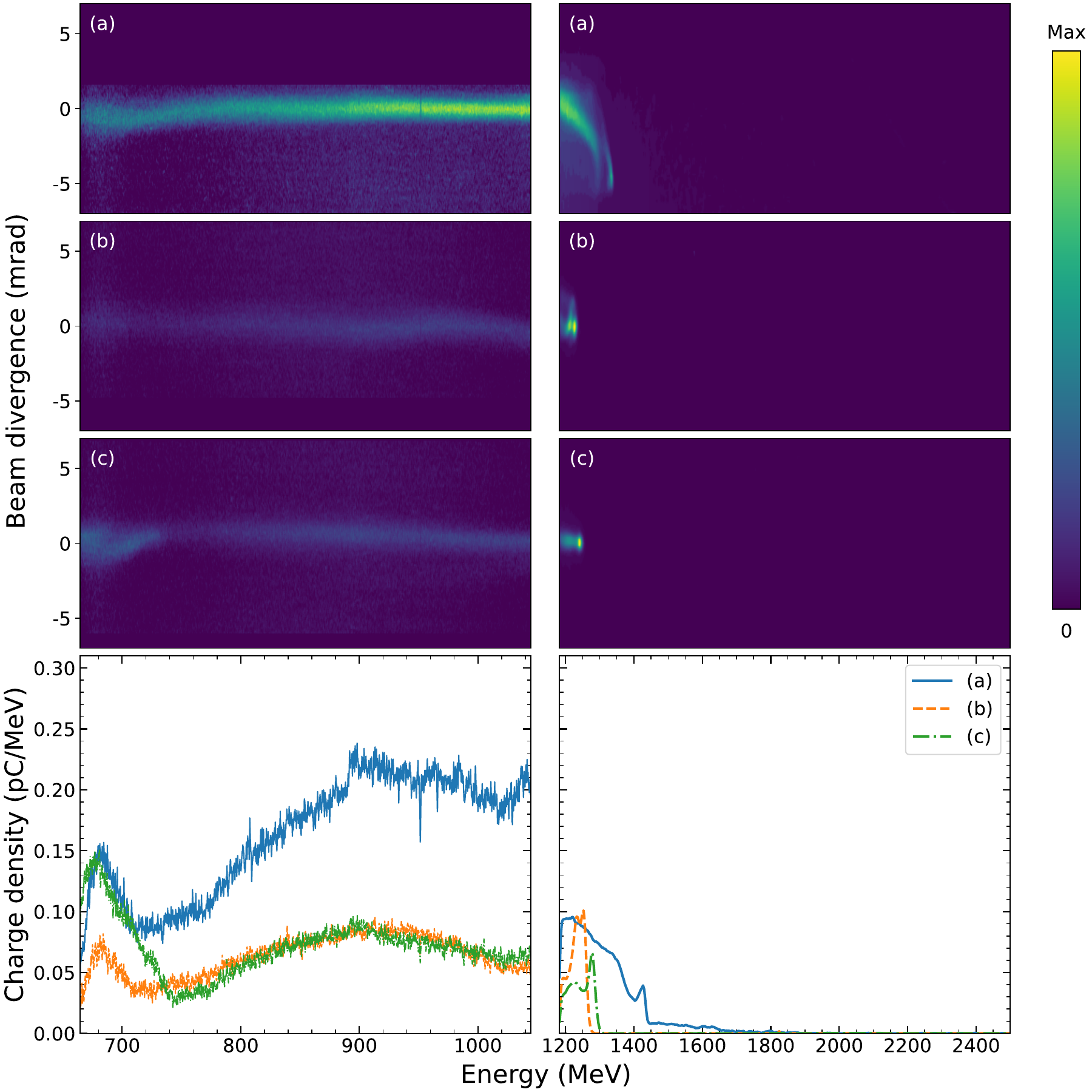}
    \caption{Top: Angularly resolved electron spectra obtained without guiding and with ionization injection. Shots in (b) and (c) are consecutive shots while (a) was obtained on a different day.  Bottom: Same spectra but angularly integrated.}
    \label{fig:pas_guide_pas_choc}
\end{figure}

With a backing pressure of 60 bar and without the guiding pulse and without the blade to trigger the injection, we obtained continuous spectra like those displayed in Fig.~\ref{fig:pas_guide_pas_choc}. The broad spectra are typical of ionization injection, where electron trapping occurs once the normalized potential vector $a_0\gtrsim~1.8$, a condition generally sustained over long distances in a laser-plasma accelerator~\cite{2006JAP....99e6109C,PhysRevLett.104.025003}. The electron beam also demonstrates good shot-to-shot stability, as illustrated by consecutive shots (b) and (c). This characteristic is inherent in the longitudinal nature of the injection technique~\cite{2013NatCo...4.1501C,2017LSA.....6E7086D}. Although the beam shows little variation from shot to shot, the day-to-day fluctuations in laser properties can lead to more significant changes, with, for instance, the maximum energy ranging from 1.25 to 1.4 GeV between days, as illustrated in Fig.~\ref{fig:pas_guide_pas_choc}.  Significant vignetting in our imaging system prevented reconstruction of the electron spectrum in the 1.05 to 1.18 GeV range. Therefore, to estimate the total charge above 750 MeV, we interpolated the signal within this range, resulting in a charge of approximately 100~pC on day one (Fig.~\ref{fig:pas_guide_pas_choc}.a) and 30~pC on day two (Fig.~\ref{fig:pas_guide_pas_choc}.b-c). The maximum charge density above 1.2 GeV is about 0.1 pC/MeV for all shots.



\begin{figure}
    \centering
    \includegraphics[width=\linewidth]{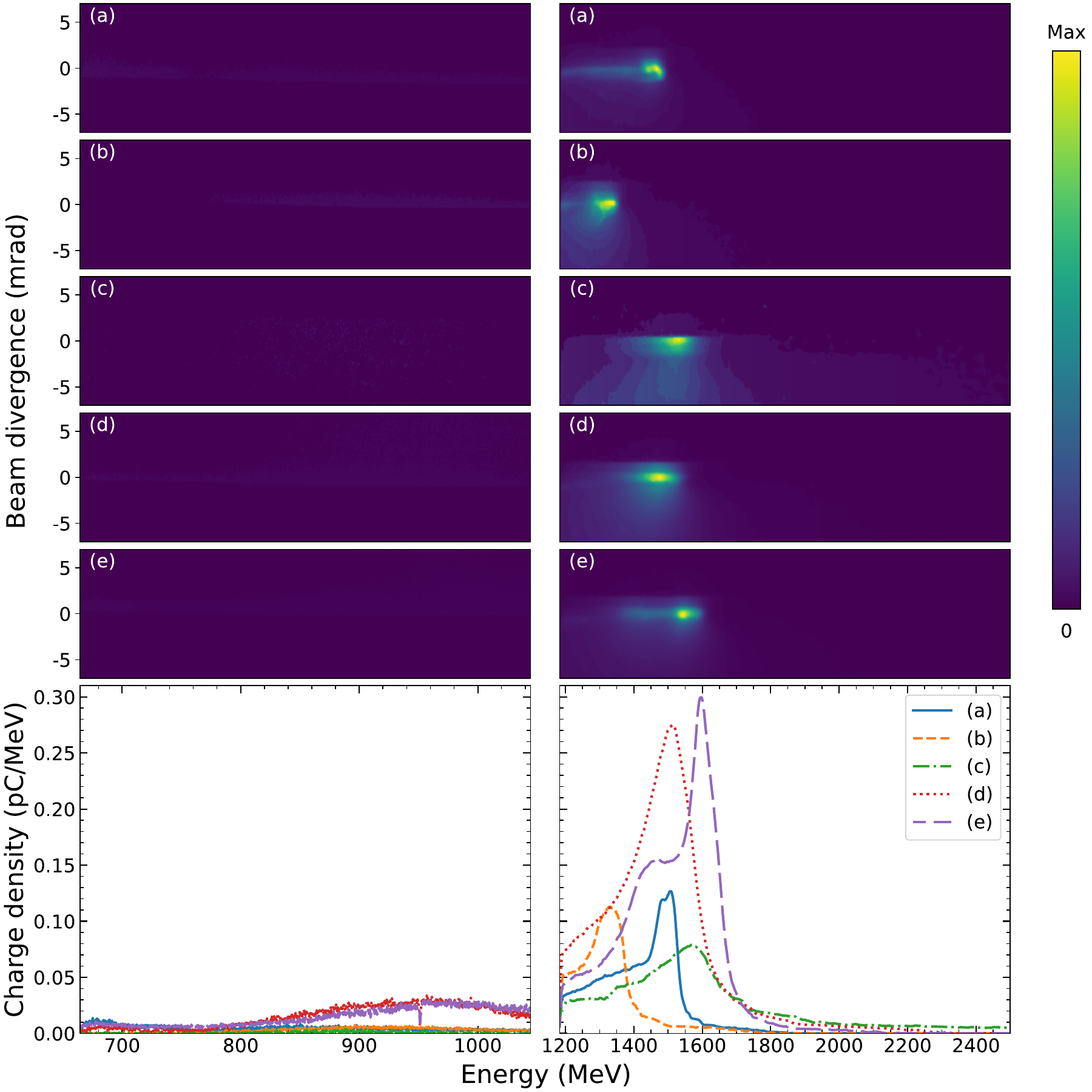}
    \caption{Top: Angularly resolved spectra obtained without guiding and with shock-assisted ionization injection. Bottom: Same spectra but angularly integrated.}
    \label{fig:pas_guide_choc}
\end{figure}

 To increase this value and achieve a peaked energy spectrum, it is necessary to localize the injection. We achieved this level of control using the shock-assisted ionization injection technique. The highest energies were achieved for a backing pressure of 50 bar, as illustrated by the five consecutive spectra shown in Fig.~\ref{fig:pas_guide_choc}. We observe high-energy peaked spectra with almost no charge below 1.1 GeV, a notably increased charge density at high energies, and a maximum energy approximately 16\% higher than without shock, exceeding 1.6 GeV for the best shots. The beam charge varies between 20.5 and 83~pC, while the relative energy spread ranges from 6.1\% to 20\% FWHM. This shot-to-shot  variation of the beam features is probably due to the poor stability of the laser focal spot. Despite this, shock-assisted ionization injection provides remarkable beam quality and stability for a PW-class laser system.



\begin{figure}
    \centering
    \includegraphics[width=\linewidth]{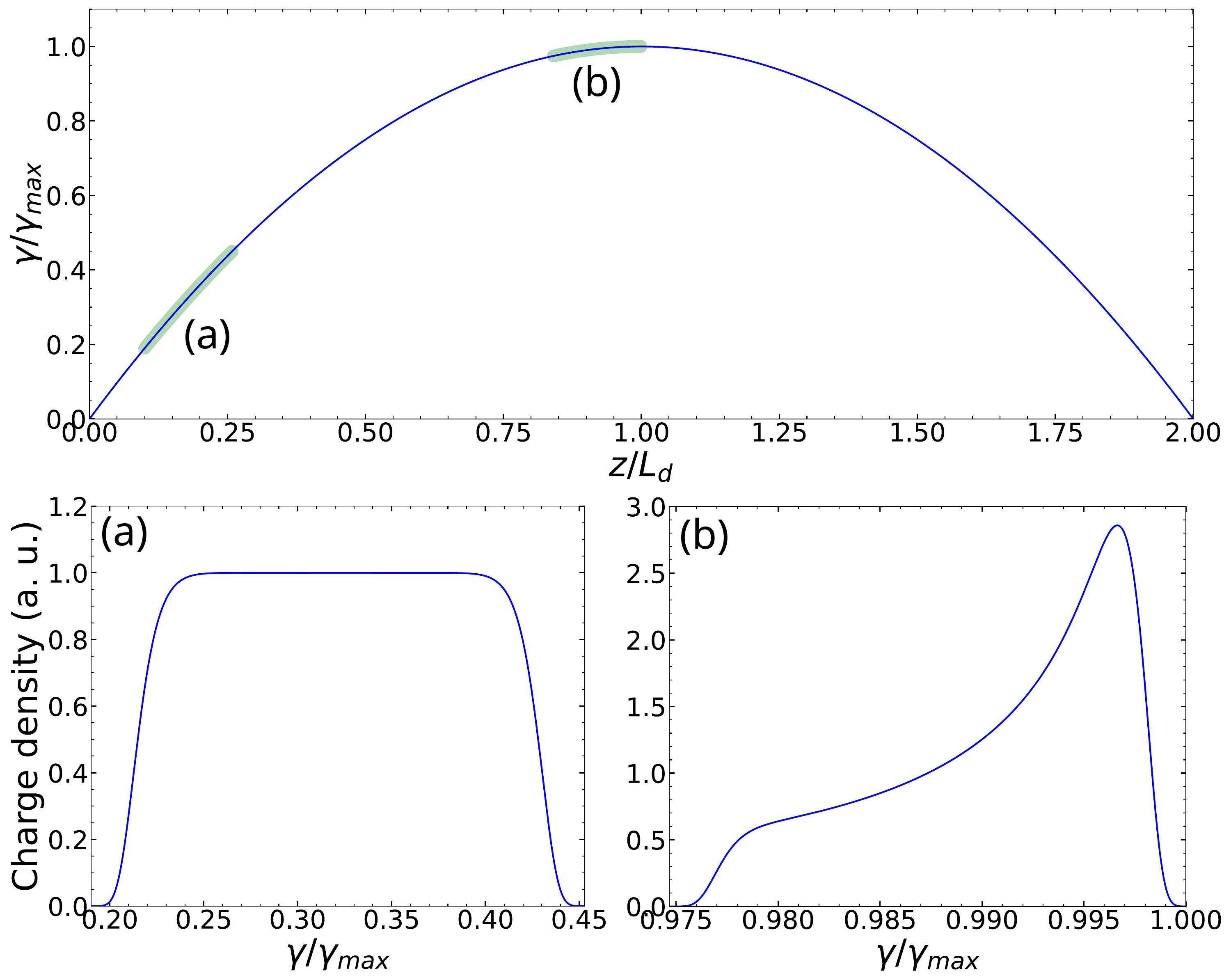}
    \caption{Electron dephasing in the bubble regime. Top panel : electrons injected at the back of the plasma cavity are accelerated to the dephasing length $L_d$ and then lose energy. The beam chirp is reduced when the beam reaches the vertex of $\gamma(z)$. Panel~(a) shows the initial beam energy spread, panel~(b) shows the steepening and decrease of the energy spread at the dephasing peak.}
    \label{fig:deph}
\end{figure}

We also observe in Fig.~\ref{fig:pas_guide_choc} that the spectra are strongly asymmetric, with a sharp drop in charge density at high energies and a slower decay towards low energies. Quantitatively, this translates into a peak-to-average energy ratio of the order of 1.1 for shots (a) and (e), which would be 1 for a symmetric beam. This suggests that the acceleration length is close to the dephasing length $L_d$. Indeed, in the bubble regime, the electrons experience a longitudinal accelerating field proportional to their position within the plasma cavity, ranging from maximum acceleration at the back of the bubble to the decelerating fields at the front. This results in a accumulation of electrons at the maximum energy as the beam reaches the center of the cavity, where the field turns negative~\cite{PhysRevLett.121.074802}. 
This phenomenon is illustrated in Fig.~\ref{fig:deph}, which depicts the ideal case of a finite-length beam injected into a linear field propagating at a speed slower than the electron bunch. As shown in panels~(a) and~(b) of Fig.~\ref{fig:deph}, while the beam's length remains constant, the energy spectrum's width is significantly compressed when it reaches the center of the cavity ($z=L_d$), with a bunching at the peak energy.
The asymmetric shape of the spectra is thus an indication that the plasma density is optimal, with the dephasing length closely matching the laser's effective propagation length. A lower plasma density would reduce the accelerating field, while a higher density would enhance dephasing, leading to reduced energy in both cases. 

We can use Lu's model to evaluate the effective accelerating length, identifying it with the dephasing length. According to this model, the maximum energy gain $\Delta W$ is linked to  the dephasing length through~\cite{PhysRevSTAB.10.061301}
\begin{align}L _d=\sqrt{\frac{3}{2}}\frac{\lambda_0}{a_0\pi} \left(\frac{\Delta W}{m_ec^2}\right)^{3/2}\mathrm{,}
\label{eq1}
\end{align}
leading in our case, for $a_0\approx 2$, to $L_d\approx 3$ cm, and a density $n_e\approx7\times10^{17}~\text{cm}^{-3}$, which is close to the measured average density, after accounting for the change in operating pressure between the measurement and the shooting conditions. It suggests that only about half of the target is used for acceleration and points out the need to guide the beam to reach higher energies. 

\begin{figure}
    \centering
    \includegraphics[width=\linewidth]{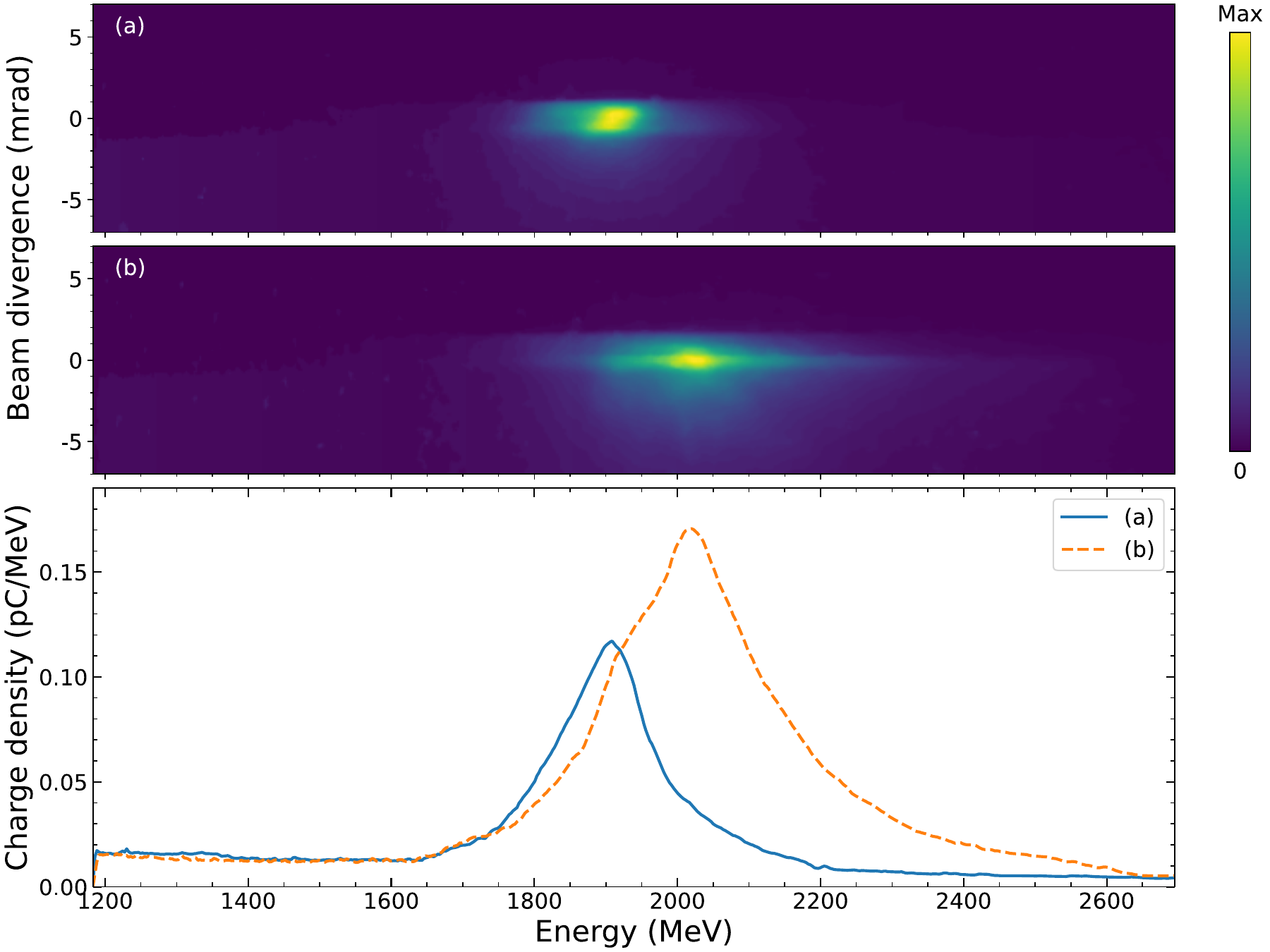}
    \caption{Top: Angularly resolved spectra obtained with guiding and shock-assisted ionization injection. Bottom: Same spectra but angularly integrated.}
    \label{fig:guide_choc}
\end{figure}

The implementation of an OFI channel allowed us to achieve this guiding. As the channel formation significantly reduces the on-axis density, we had to increase the backing pressure to keep the density high enough for injection. With a backing pressure of 70 bar, the maximum available, we obtained the spectra displayed in Fig.~\ref{fig:guide_choc}. These spectra were recorded immediately after those shown in Fig~~\ref{fig:pas_guide_choc}, under the same laser conditions.    Due to the pointing instability of the main laser, the coupling efficiency into the guide fluctuated significantly from shot to shot, resulting in many shots with little or no charge injection~\cite{Oubrerie2022}. Figure~\ref{fig:guide_choc}, therefore, displays only the two best shots from a series of 18 shots. Despite being selected spectra, it is worth noting that the maximum energy achieved in these shots significantly surpasses that measured in all shots without guiding.
The spectra show relative energy spreads (9\% and 12\%, respectively) and beam charges (21 and 47 pC) comparable to those measured without the waveguide, while achieving significantly higher energies (1.9 and 2~GeV). This thus demonstrates a successful increase in energy gain by $400–500$ MeV, achieved through guiding, without any loss in charge or beam quality.




\section{Discussion}

\begin{figure}
    \centering
    \includegraphics[width=\linewidth]{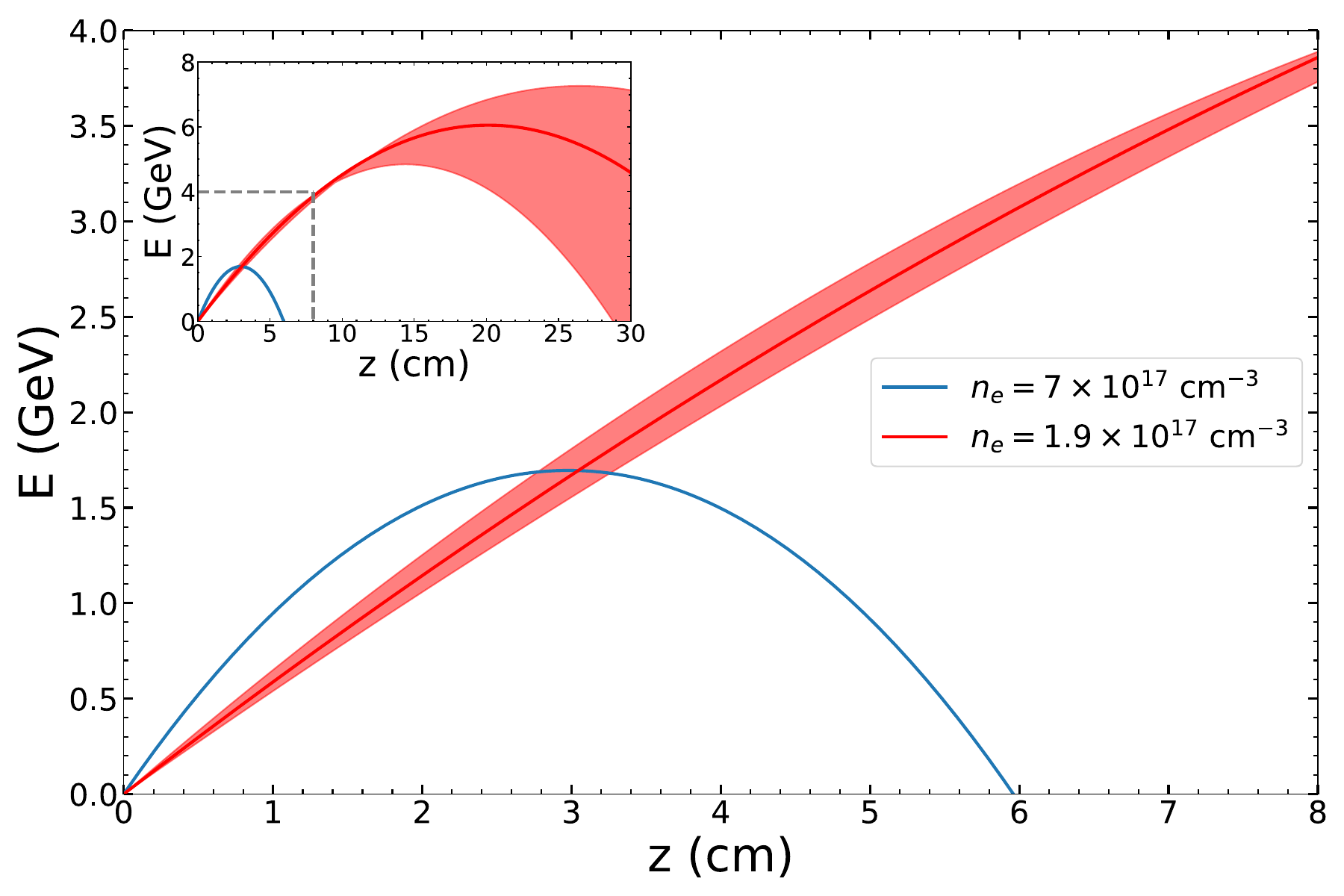}
    \caption{Gain of a laser plasma accelerator depending on the electronic density. The blue curve corresponds to the density in the non-guided case. The red shaded area delimits the energy gain for a density in the guiding channel reduced by a factor of four and by a factor of six compared to the density without the guiding beam, and the red curve corresponds to a factor of five. The inset corresponds to the energy gain over 30~cm, while the bigger plot represents the same gain but over 8~cm.}
    \label{fig:comp_gain}
\end{figure}

To assess the performance of the accelerator, we have to estimate the density in the guiding channel. From previous studies~\cite{PhysRevE.97.053203}, we can estimate that the channel-forming beam reduces the on-axis density by a factor of 4 to 6. Taking into account the increase in operating pressure between the non-guided case and the guided case (from 50 to 70~bar) leads to a density in the guiding channel between 2 and 3$\times10^{17}~\text{cm}^{-3}$. Using Lu's model and assuming a parabolic evolution of the energy gain due to dephasing, we can estimate the optimal acceleration length and maximum energy gain for these densities. For a deep channel, the maximum energy gain is around 7.3~GeV over 26~cm of acceleration, and a shallow channel leads to around 4.8~GeV of energy gain over 14~cm, as shown in the inset of Fig.~\ref{fig:comp_gain}. Considering smaller acceleration lengths, closer to the length of our target, we estimate the energy gain over 6~cm of acceleration with a constant density profile to be between 2.9 and 3.2~GeV, as shown in Fig.~\ref{fig:comp_gain}, which is slightly higher than the cutoff energy measured with the guiding beam (Fig.~\ref{fig:guide_choc}).  This discrepancy could be attributed to potential inhomogeneities in the target density profile, particularly a descending density gradient at the end of the propagation. Such a gradient reduces the amplitude of the accelerating electric field and increases the dephasing rate.

In summary, we demonstrated the production of a multi-GeV electron beam with a peaked energy spectrum using a hydrodynamic shock for controlled injection and an OFI waveguide to guide the laser through a 6 cm-long plasma. The results suggest that the electron energy is primarily limited by the target length. With the same plasma density and laser energy as in Fig.\ref{fig:guide_choc}, extending the target to tens of centimeters could enable energy gains exceeding 5 GeV, similar to those achieved at other PW-laser facilities~\cite{PhysRevLett.125.074801,10.1063/5.0097214}. To further optimize the setup, the implementation of an active pointing stabilization system~\cite{10.1063/1.3556438} could ensure a precise overlap of the guiding and driver beams, improving laser coupling into the waveguide on every shot. This would enhance both the stability and quality of the electron beam, leading to a reliable multi-GeV source suitable for demanding applications such as free-electron lasers~\cite{2021Natur.595..516W,2023NaPho..17..150L} or the probing of quantum electrodynamic processes~\cite{2012RvMP...84.1177D}.

\begin{acknowledgments}
The authors acknowledge the national research infrastructure Apollon and the LULI for their technical assistance. The project has received funding from the European Union’s Horizon 2020 Research and Innovation programme under the grant agreement no. 871124, iFAST and from  the French Agence Nationale de la Recherche (ANR) under
project ANR-19-TERC-0001-01.
\end{acknowledgments}

\bibliographystyle{apsrev4-2}
\bibliography{bib}

\end{document}